# Jupiter – friend or foe? III: the Oort cloud comets

## J. Horner[1], B. W. Jones[1] & J. Chambers[2]

[1]*Astronomy Group, Physics & Astronomy, The Open University, Milton Keynes, MK7 6AA, UK*

[2]*Carnegie Institution of Washington, 5241 Broad Branch Road NW, Washington DC, 20015, USA*

e-mail: b.w.jones@open.ac.uk    Phone: +44 (0) 1908 653229    Fax: +44 (0) 1908 654192

**(SHORT TITLE: Jupiter – friend or foe? III: the Oort cloud comets)**

**Abstract**

It has long been assumed that the planet Jupiter acts as a giant shield, significantly lowering the impact rate of minor bodies on the Earth. However, until recently, very little work had been carried out examining the role played by Jupiter in determining the frequency of such collisions. In this work, the third of a series of papers, we examine the degree to which the impact rate on Earth resulting from the Oort cloud comets is enhanced or lessened by the presence of a giant planet in a Jupiter-like orbit, in an attempt to more fully understand the impact regime under which life on Earth has developed. Our results show that the presence of a giant planet in a Jupiter-like orbit significantly alters the impact rate of Oort cloud comets on the Earth, decreasing the rate as the mass of the giant planet increases. The greatest bombardment flux is observed when no giant planet is present.

**Key words:** Solar System – general, comets – general, Oort cloud comets, long-period comets, minor planets, asteroids, planets and satellites – general, Solar System – formation.

**Introduction**

In our previous two papers "Jupiter – friend or foe? I: the Asteroids" (Horner & Jones, 2008, Paper I), and "Jupiter – friend of foe? II: the Centaurs" (Horner & Jones 2009, Paper II), we pointed out that it is widely accepted in the scientific community (and beyond) that Jupiter has significantly reduced the impact rate of minor bodies on the Earth, notably small asteroids and comets. As well as causing local mayhem in the biosphere, larger impacts can surely result in mass extinctions, and will therefore have had a major influence on the survival and evolution of life (Alvarez et al., 1980, Sleep et al., 1989). However, the effects of such impacts are not solely damaging to the development of advanced life – indeed, without extinctions, far fewer empty ecological niches would appear to promote the emergence of new species. Alternatively, really large impacts could occur so often that the evolution of a biosphere would be stunted by overly frequent mass extinctions, each bordering on (or resulting in) global sterilisation. Without Jupiter present, it has been argued, such frequent mass extinctions would occur (Ward & Brownlee 2000).

It is perhaps surprising, when one considers how widely this well established view of Jupiter's protective role is held, that very little work has been carried out to back up that hypothesis. Indeed, until recently, almost no studies have examined the effects of the giant planets on the flux of minor bodies through the inner Solar System. Wetherill (1994) showed that, in systems containing giant planets which grew only to the mass of around Uranus and Neptune, the impact flux of cometary bodies, experienced by any terrestrial planet, would be a factor of a thousand times greater than that seen today.

There are two reservoirs of cometary bodies. One is the Oort cloud, a predominantly spherical distribution of $10^{12}$-$10^{13}$ icy bodies, the great majority of which are smaller than 10 km in diameter, occupying a thick shell ranging from approximately $10^3$-$10^5$ AU from the Sun (e.g. Horner & Evans, 2002). Objects perturbed inwards from this cloud become the long period comets (periods >~ 200 years, with the full range of orbital inclinations). The other is the Edgeworth-Kuiper belt, a population of icy-rocky bodies, again predominantly less than a few tens of km across[1], orbiting beyond Neptune in fairly low inclination orbits. Associated with the Edgeworth-Kuiper belt is a less stable, more dynamically excited component, known as the Scattered Disk (see e.g. Lykawka & Mukai, 2007; Gomes et al. 2008 ). The orbits of objects within the Scattered Disk are typically somewhat unstable, and it is thought that a steady trickle of objects evolve inwards from this belt to eventually become the short period comets.

Wetherill obtains his 1000 or so factor by applying a Monte-Carlo simulation to a population of bodies initially occupying eccentric, low inclination orbits with semimajor axes in the range 5-75 AU. Jupiter orbits at 5.2 AU so this population is bound to be far more sensitive to the mass of Jupiter and Saturn than bodies derived from the Edgeworth-Kuiper belt, thus exaggerating greatly the shielding provided. Our Paper II is consistent with this. Less importantly, he employed a Monte-Carlo simulation which would have yielded numerical data less reliable than modern day orbital integrators, and at a time when far slower computers placed much greater practical limits on the range of parameters that could be studied.

Nevertheless, Wetherill's results were very convincing, and for a decade, no more work was done to examine this subject. In more recent times (see Paper I), a study by Laasko et al. (2006) led to the conclusion that Jupiter "*in its current orbit, may provide a minimal of protection to the Earth*". They also mention the work of Gomes et al. (2005), from which it is clear that removing Jupiter from our Solar System would result in far fewer impacts on the Earth by lessening or removing entirely the effects of the proposed Late Heavy Bombardment of the inner Solar System, some 700

---

[1] The objects currently known in the Edgeworth-Kuiper belt range in size to over 2000 km in diameter. However, large objects are over-represented because they are easier to discover. A whimsical analogy is on the plains of Africa – even though there are billions of flies within a few kilometres, it's far easier to see the few elephants over such a distance.

Myr after its formation, but nothing is said about more recent times.

The idea that Jupiter has protected the Earth from excessive bombardment dates back to when the main impact risk to the Earth was thought to arise from the Oort cloud comets. The idea probably dates back to the 1960s, when craters were first widely accepted as evidence of ongoing impacts upon the Earth, and far more long period comets were known than the combined numbers of short period comets and near-Earth asteroids combined. It is well known that a large fraction of such objects are expelled from the Solar System after their first pass through the inner Solar System, mainly as a result of Jovian perturbations. Hence, by significantly reducing the population of returning objects, Jupiter lowers the chance of one of these cosmic bullets striking the Earth. However, in recent years, it has become accepted that near-Earth objects (some of which come from the asteroid belt, others from the short period comet population) pose a far greater threat to the Earth than that posed by the Oort cloud comets. Indeed, it has been suggested that the total cometary contribution to the impact hazard may be no higher than about a quarter (e.g. Chapman & Morrison, 1994, Morbidelli et al. 2002). The effect of Jupiter on the three immediate source populations of potentially hazardous objects, asteroids, the Centaurs (see below), and the Oort cloud comets, has been neglected, and in order to ascertain the overall effect of Jupiter on the terrestrial impact flux, it is important to understand its influence on each of the three kinds of threatening body.

In Paper I we examined the effect of changing Jupiter's mass on the impact rate experienced by the Earth from objects flung inwards from the asteroid belt. We performed numerical integrations for a simulated time of 10 Myr, with a mass range from 1% of Jupiter's mass ($M_J$), to twice its mass. Our results were surprising. Table 1 in Paper I shows that at 1.00 $M_J$, the number of impacts on the Earth is about 3.5 times the number at 0.01 $M_J$ – hardly a shield! Between these two "Jupiter" masses there is a peak at around 0.2 $M_J$ where the number of impacts is nearly double that at 1.00 $M_J$. We'll return to this result in the discussion.

In Paper II, the Centaurs constituted the source of potential bombarders. These objects originate in the trans-Neptunian region, and have evolved onto dynamically unstable, planet crossing orbits in the outer Solar System. They represent the direct parent population of the short period comets, with previous studies (e.g. Horner et al., 2004) suggesting that over 30% of all Centaurs will become short period comets at some point in their lifetime. Known Centaurs in our own planetary system were used to create a population of over 100,000 bodies, initially located well beyond the gravitational influence of Jupiter (which extends to about 6.3 AU (three Hill radii from Jupiter). Indeed, to ensure that the population chosen had not recently been influenced by the giant planet, no object was selected that had a perihelion distance closer to the Sun than Uranus. The evolution of these test particles was again followed in Solar Systems with Jupiters of various mass, for a period of 10 Myr. The mass of the Jupiters studied in Paper II ranged from zero to twice Jupiter's mass. Table 1 in that paper shows that at 1.00 $M_J$ the number of impacts on the Earth is about the same as the number at zero mass and 0.01 $M_J$. Between these extremes there is again a peak at around 0.2 $M_J$ where the number of impacts is 4.5 times that at 1.00 $M_J$. We'll also return to this result in the discussion.

The results from Paper I and II show conclusively that the idea of "Jupiter – the shield" is far from a complete description of how giant planets affect terrestrial impact fluxes, and that more work is needed to examine the problem.

In this paper, we detail the results of simulations examining the role of Jupiter in modifying the impact risk due to the long period comets, which come from the Oort cloud. Historically, any comet with a period greater than 200 years was considered a "long-period comets", although those on their first pass through the inner Solar System typically have orbital periods over $10^5$ years. These "new" long period comets are sent into the inner Solar System as a result of distant gravitational perturbations from passing stars, passing dense molecular clouds, and by the Galactic tide (Emelyanenko et al., 2007, Nurmi et al. 2001).

**Simulations**

In order to create a swarm of test objects which might evolve onto Earth-impacting orbits, we randomly generated a population of 100,000 massless test particles, with perihelia located in the range 0.1 - 10 AU and aphelia between $10^4$ and $10^5$ AU. The population was structured in an attempt to emulate the observed aphelion distribution of long period comets, with a peak at around 40,000 AU (corresponding to a semi-major axis of 20,000 AU). Thus, the median aphelion distance was set at 40,000 AU, with the probability of a test particle having a value higher or lower than that value falling linearly to 0 at the boundaries. Due to the skewed distribution produced, this leads to a mean aphelion distance of 50,000 AU for the sample. This distribution is a simple, but effective, attempt to fit the known distribution of new Oort cloud comets (see e.g. Horner & Evans, 2002, and references therein). The likelihood of an object having a given perihelion distance $q$ was calculated so that the majority of the comets had larger initial perihelion values. Therefore, the value of $q$ was

determined as follows

$$q = 0.1 + ((q_{max} - q_{min})^{3/2} \times random)^{2/3}$$

where $q_{max}$ and $q_{min}$ are the maximum and minimum possible perihelion distances of 0.1 and 10 AU, respectively, and *random* is a random number between 0 and 1, generated within the cloning program, such that the initial distribution of *q* is as in Figure 1. This resulted in approximately 3% of the initial sample having orbits that cross the Earth's orbit (Earth-crossing orbits), and approximately 38% being on initially Jupiter-crossing orbits (orbits with *q* less than, or equal to, 5.203).

> Figure 1  (Figures are after the text) The initial perihelion and semi-major axis distributions of our test particles. The upper panels show the cumulative distributions as a function of semi-major axis (left) and perihelion distance (right), while the lower panels show the binned distributions as a function of these values, split in such a way that 1000 bins cover the entire spread of possible values. The noise apparent in the lower panels is the result of the random number generator used. As can be seen in the upper panels, when the number of particles is high enough, the distributions become smooth.

The inclination of a comet's orbit was set randomly between 0 and 180 degrees, the longitude of the ascending node and the argument of perihelion were each set randomly between 0 and 360 degrees. Finally, the location of the comet on its orbit, at the start of the integration (the initial mean anomaly) was set randomly between 0 and 360 degrees. Again, the distributions tend to be uniform as the number of test particles increases.

Once the cloning process was complete, 100,000 test particles were distributed on a wide variety of long period orbits. The dynamical evolution of these massless particles was then followed for a period of 100 million years using the hybrid integrator contained within a version of the *MERCURY* (Chambers, 1999) package, that had been modified in order to allow orbits to be followed in barycentric, rather than heliocentric, coordinates. The evolution of the particles' orbits occurred under the influence of the planets Jupiter and Saturn, both of which had initial orbital elements equal to their present values (though they barely changed during the simulation), and were treated as fully interacting massive particles. The integration length, time-step (180 days), and the number of planets included, were chosen to provide a balance between required computation time and the statistical significance of the results obtained. In the simulation the cometary bodies were treated as massless particles, and were unable to gravitationally interact with each other. They felt the gravitational influence of the Sun, Jupiter and Saturn, but did not exert any force on those bodies.

Whereas in Papers I and II we counted the number of collisions on an (inflated) Earth, for the Oort cloud comets a different approach was needed. The orbital period of Oort cloud comets is so great that, even in a 100 Myr simulation, very few close encounters with the Earth would be expected even were the Earth greatly inflated. Therefore, in order to directly acquire the rate of impacts on the Earth, we would have had to simulate a vast number of test particles, many orders of magnitude higher than that used. This, in turn, would have required an infeasibly large amount of computation time, and so a different method for calculating the resulting impact flux was required.

A proxy for the impact rate was clearly needed, and we initially chose to use the number of comets that survived as the orbital integration proceeded. Over the course of the integrations, comets were followed as they moved around the Sun until they hit Jupiter, Saturn, or the Sun, or were ejected from the Solar System entirely. Since comets thrown to sufficiently large distances will clearly never return (even if their eccentricity is slightly less than one), due to the un-modelled gravitational effects of nearby stars, the galactic tide, and molecular clouds, the particles in our simulations were considered "ejected" when they reached a barycentric distance of 200,000 AU - twice the maximum initial aphelion distance. Note that our work focuses on comets after they have been sent inwards, so the fate of departing survivors beyond 200,000 AU is not of importance in our work.

As the comets in our simulations orbit around the Sun, they suffered orbital perturbations around the time of perihelion passage that resulted from the distant influence of Jupiter and Saturn. These perturbations act to either lengthen or reduce the orbital period of the comet in a random manner. However, the comets are so loosely bound to the Sun that only a moderate change in their orbital angular momentum is sufficient to remove them from the system entirely. Clearly, a comet whose orbital period is reduced, such as C/1995 O1 Hale-Bopp, a comet that most likely originated in the Oort Cloud, but then captured onto a much shorter period orbit (~2500 years for its next trip around the Sun) due to the ongoing effect of the giant planets, will return to potentially threaten the Earth, while one that is ejected from the system can never return to pose a threat. It is therefore clear that, for a given population, the greater the number of objects which survive, the higher the impact rate experienced by the Earth.

Non-gravitational forces (such as those that would result from jetting or splitting of the cometary nucleus) were neglected, and no perturbations were applied to the comets to simulate the effect of passing stars, the galactic tide, and passing molecular clouds. Although this means that our simulations are a simplication of the true situation, the effect of these distant perturbations would be the same for all masses of Jupiter, and so they can safely be neglected.

The mass of "Jupiter" used in our simulations was modified so that we ran five separate scenarios. Planets with 0.00, 0.25, 0.50, 1.00, and 2.00 times the mass of the present Jupiter were used. The (initial) orbital elements of "Jupiter" were identical in all cases.

Were Jupiter a different mass, the architecture of the outer Solar System would probably be somewhat different, but rather than try to quantify the uncertain effects of a change to the configuration of our own Solar System, we felt it best to follow the same procedure as in Papers I and II, and change solely the mass of the "Jupiter", and therefore work with a known, albeit modified, system rather than a theoretical construct. For a flux of objects moving inwards from the Oort cloud, this does not seem unreasonable – by choosing a population of objects well beyond the "Jupiter" in our simulations, with initial aphelia between $10^4$ and $10^5$ AU (and considering our test particles to represent dynamically "new" comets, on their first pass through the inner Solar System), the planet's influence on the objects prior to the start of our simulations is negligible. We believe this method allows us to make a fair assessment of the role of Jovian mass on such objects.

The complete suite of integrations ran for some four months of real time, spread over the cluster of machines sited at the Open University. This span of real time equates to over thirteen years of computation time, and resulted in measures of the comet survival rate in each of the five mass scenarios.

**Results**

Figure 2 shows the number of surviving comets (comets that have not yet reached a barycentric distance of 200,000 AU, or collided with the Sun, Jupiter or Saturn) versus time for each of the five scenarios. The differences between the scenarios quickly become apparent, with the high-mass cases seeing a significantly more rapid loss of comets than those of low-mass. The lower plot of Figure 2 shows this decay in the form of a log-log plot, which reveals that this enhanced ejection rate continues to the very end of our simulations, by which point only a small fraction of the initial cometary population remains. Clearly, at some point beyond the 100 Myr of our integrations, the systems would become totally depleted in cometary bodies, and would once again be indistinguishable. However, in reality, the situation is not quite so simple. The long period cometary population is continually replenished from the Oort cloud, and so most likely exists in a steady state, with the loss of objects matched by the introduction of new ones. As such, it is clear that any scenario with a shorter mean lifetime for long period comets (a greater rate of loss) will have a significantly smaller steady state cometary population, and that population will therefore pose less of a risk to the Earth.

> Figure 2    The number of surviving Oort cloud comet clones as a function of time into the orbital integration. Panel (a) has linear scales on the axes, and the (b) logarithmic scales. Five scenarios are shown, all with "Jupiter" in Jupiter's present-day orbit. From top to bottom, the mass of the giant in Jupiter's orbit as a multiple of Jupiter's mass is 0.00, 0.25, 0.50, 1.00. and 2.00. The (initial) orbits of both "Jupiter" and Saturn (the two perturbing planets) were the same as modern day orbits in all scenarios, as were the initial orbits of the 100 000 test particles. In other words, the only difference between the five integrations was the mass of our "Jupiter".

In reality, the actual number of long-period comets on orbits that bring them closer to the Sun than 10 AU is far greater than that portrayed in our simulations. If we assume the mean semi-major axis of a typical (new) long period comet is 20,000 AU, then the mean orbital period will be of the order of three million years. Every year, at a very conservative estimate, at least ten such comets are discovered (and, given that the bulk of these comets are discovered as they pass within the orbit of Jupiter, it seems certain this is only the tip of the iceberg) – and so it is clear that, to maintain this level of new comets, there must be many millions of such objects currently on orbits that bring them closer that 10 AU from the Sun. As such, the number of comets used in this work is clearly significantly fewer than the real population. However, enough cometary bodies were used in this study that the results are statistically significant, and attempting to increase the population by another factor of ten, or even one hundred, would have led to an unacceptable increase in the time required for the integrations to become complete.

When one looks at the upper panel of Figure 2, the difference between the scenarios is most marked between ten and twenty million years after the start of the simulations. At 10 million years, for example, the number of survivors decreases steadily from the scenario in which there is no "Jupiter" present, to that when a giant planet of twice Jupiter's mass is present. At zero mass the number of survivors is a little under 60 000, which is a little under 60% of the initial 100 000 at zero time. This number decreases monotonically as "Jupiter's" mass increases, to around 32 000 (32%) at one Jupiter mass, and around 23 000 (23%) at two Jupiter masses. Table 1 presents a selection of data from Figure 2.

With no Jupiter present, Saturn (as the only remaining massive body in the integrations) must be solely responsible for ejecting the Oort cloud comets.

Table 1   The number of surviving Oort cloud comet clones at various times into the orbital integration.

| Mass/$M_J$[1] | 0 | 1.00 Myr | 10.0 Myr | 100 Myr |
|---|---|---|---|---|
| 0.00 | 100 000 | 99982 | 58949 | 3689 |
| 0.25 | 100 000 | 99861 | 50138 | 2551 |
| 0.50 | 100 000 | 99681 | 41835 | 2337 |
| 1.00 | 100 000 | 99314 | 32334 | 1495 |
| 2.00 | 100 000 | 98659 | 23253 | 852 |

(1) $M_J$ is the mass of Jupiter.

It is possible, however, that the collision rate on Earth is not simply proportional to the number of surviving Oort cloud comets. There are two additional possibilities.

1. There could be preferential survival of *either* the Oort cloud comets that cross Earth's orbit ($q < 1$ AU), *or* those that do not ($q > 1$ AU). The outcome could be sensitive to "Jupiter's" mass.
2. Likewise possibly sensitive to "Jupiter's" mass is the orbital period distribution of Oort cloud comets. Giant planets with different masses might be more or less efficient at capturing comets onto shorter period orbits. Objects captured in this manner would clearly have correspondingly higher frequencies of perihelion passages, and would constitute a greater threat than their brethren on longer period orbits. Depending on the efficiency with which such capture happens, compared to that of ejection, it could be the case that systems with fewer surviving comets (trapped on shorter period orbits) would pose a greater impact threat than those with more comets, trapped on longer orbits.

To explore the first possibility, we plotted, for the various "Jupiter" masses, the log of the number of surviving Oort cloud comets as a function of time for objects with $q < 1$ (Earth-crossing), $q < 1.524$ AU (Mars-crossing), and $q < 5.203$ AU (Jupiter-crossing), along with the total number of survivors (for reference). The outcome is shown in Figure 3. It is immediately apparent that comets on Jupiter-crossing orbits are more efficiently ejected than those with perihelia beyond the giant planet, which is just as expected. It is also apparent that, for all three values of maximum perihelia, that the number of survivors falls more rapidly as the mass of our "Jupiter" is increased – there is no preferential survival of comets on Earth, or Mars crossing orbits, though there is preferential ejection of comets with their perihelia closer to the Sun than the orbit of Jupiter.

> Figure 3    The number of surviving Oort cloud comet clones as a function of time into the orbital integration, for a variety of maximum perihelion values. Our five scenarios are again shown with the different Jovian masses coloured as follows – 0.00 $M_J$, 0.25 $M_J$, 0.50 $M_J$, 1.00 $M_J$ and 2.00 $M_J$. The only difference between the five integrations was the mass of the "Jupiter" used – everything else was kept constant. The upper left-hand plot shows the log of the number of surviving Oort cloud comets against time, with the plot beneath (lower left-hand panel) showing the log of the number of comets that survive on Jupiter-crossing orbits (i.e. any orbit with a perihelion distance less than the Jovian semi-major axis), again as a function of time. The two right hand panels show the log of the number of surviving comets that lie on Earth-crossing (top) and Mars-crossing (bottom) orbits against time. Note that, even though the decay in Earth and Mars crossing objects is so rapid that the number quickly falls into the realm of small number statistics; the same trend is visible in all four plots – the more massive the "Jupiter" in that system, the more rapidly the comets are ejected.

As can be seen in the top right panel of Figure 3, the number of objects on Earth-crossing orbits falls away rapidly with increasing time, even when no Jupiter is present. Indeed, the number remaining plummets into the realm of small number statistics after only around 10 million years. Nevertheless, it is clear that, as a first order approximation, the behaviour of the number of surviving Earth-crossing objects is strongly similar to that of the Jupiter-crossers – again, not entirely unexpected, since the two drivers of the orbital evolution of these comets (Jupiter and Saturn) lie at distances beyond the perihelia of these objects.

To investigate the second possibility, the possible sensitivity to "Jupiter's" mass of the orbital period distribution of Oort cloud comets, we examined the behaviour of Jupiter-crossing objects as a function of time, using them as a proxy for the smaller Earth-crossing population. In the upper panel of Figure 4, for reference, we show the log of the number of surviving Jupiter-crossers as a function of time and "Jupiter" mass. In the second panel, we plot the log of the mean orbital period (in years) of those comets as a function of time (studying only those moving on bound orbits, and ignoring any that were parabolic or hyperbolic). The third panel shows the mean of the inverse orbital periods (calculated by obtaining the inverse orbital period for each object, summing them together, then dividing by the number of objects considered). The bottom panel of Figure 4 shows the evolution of the log of a simplified estimate of the collision probability resulting from these comets, as a function of time. This estimated collision probability, *PCol*, was simply obtained by multiplying the number of surviving objects by the mean of their inverse orbital periods – it effectively measures the frequency of perihelion passages by the comets in question, to an arbitrary scale. It is obvious that, for the Jupiter-crossing comets, the probability of collision (the frequency of perihelion passages) falls away dramatically as a function of time, with the greatest and most rapid falls occurring for those scenarios in which the Jupiter is most massive. Given that the mass of "Jupiter" has only a slight effect on the mean orbital period (middle panel), it is the increased efficiency with which Oort cloud comets are ejected from the Solar System that results in a decrease in the collision probability as the mass of "Jupiter" increases.

> Figure 4    In the upper panel, the log of the number of surviving comets on Jupiter-crossing orbits is plotted as a function of time through the integration. The second panel shows the log of the mean orbital period for all surviving Jupiter-crossers, again as a function of time. The third panel shows the mean of the inverse orbital periods (calculated by obtaining the inverse orbital period for each object, summing them together, then dividing by the number of objects considered). The lowest panel shows the log of a simple estimate of the collision probability resulting from those comets with any given planet. This was obtained by merely multiplying the mean of the inverse orbital periods (panel 3) with the number of objects remaining on a Jupiter-crossing orbit – effectively it can be considered a relative "encounter-frequency", which is clearly directly related to the impact rate. As before, the different Jovian masses are coloured as follows – 0.00 $M_J$, 0.25 $M_J$, 0.50 $M_J$, 1.00 $M_J$ and 2.00 $M_J$.

**Discussion**

From Figure 2 and Table 1 it is clear that a giant planet in Jupiter's orbit does provide a measure of protection to planets in the inner Solar System from bombardment by Oort cloud comets.

In Papers I and II we report the outcome of orbital integrations where, respectively, the potential bombarders are the asteroids and the Centaurs[2]. As pointed out in the Introduction, for the asteroids the impact rate for systems with a one Jupiter-mass "Jupiter" present is about 3.5 times greater than if that planet were just 1% of Jupiter's mass. For the Centaurs, the factor is close to one. These two populations account for most of the impacts on the Earth, with a proportion of only a quarter or so for the contribution of all comets to the terrestrial impact flux (e.g. Chapman & Morrison, 1994, Morbidelli et al. 2002). Note that the asteroids and Centaurs have low inclination orbits which increases the collision probability *per object*, whereas the Oort cloud comets come from all inclinations, which effectively reduces their comparative collision probability. However, the typical collision velocity will generally be much higher for an Oort cloud comet than an object originating in one of the other two populations. This acts to increase the relative importance of the Oort cloud comets as bombarders. Despite this, at the current time, they remain as minor players in the bombardment of the Earth.

---

[2] The Centaurs are objects moving on dynamically unstable orbits among the giant planets, and are the direct parent population of the short period comets. They themselves are sourced primarily from the Scattered Disk, although objects leaving the Neptunian Trojan cloud may also make a significant contribution (Horner & Lykawka, 2009). A significant fraction (Horner et al. 2004) of the Centaurs evolve to become short period comets over their lifetimes, keeping the population at a roughly constant level.

The outcome of our work, taken as a whole, shows that Jupiter has not protected the Earth from bombardment – in fact, it seems far better to have no massive planet in Jupiter's orbit. However, a one Jupiter-mass planet is significantly less threatening, for the cases of Centaurs and asteroids, than one around the mass of the planet Saturn, which leads to a potentially catastrophic increase in the impact flux that would be experienced by the Earth.

Why is there a peak in the impact rate for both the asteroids and the Centaurs? For the Centaurs (the source of the short-period comets), the explanation is actually remarkably simple, and is detailed in Paper II. In brief, it arises from the balance between "Jupiter" placing small bodies on Earth-crossing orbits and "Jupiter" ejecting them from the Solar System, a balance that depends on "Jupiter's" mass.

For the asteroids, the situation is a little more complex; see Paper I. In short, the peak is primarily a result of the changes in depth, breadth and location of the $\nu_6$ secular resonance in the main asteroid belt as the mass of "Jupiter" is changed. When "Jupiter" is around the mass of Saturn, this deep, destabilising resonance lies in the middle of the asteroid belt, and acts to stir up a huge region that would otherwise be stable. This great region of instability leads to a greatly increased flux of asteroids to the inner Solar System, and in turn causes a significant increase in the impact rate experienced by the Earth.

The case for the long period comets is significantly different from the two scenarios just discussed. Objects inbound from the Oort cloud are only tenuously bound to the Solar System, and it only takes a remarkably small perturbation to act to modify their orbit in such a way that, once they leave its inner reaches, they will never return. As such, even the most distant encounters between a comet and a planet can act to remove it from the system for ever. Clearly, the more long period comets are ejected from the Solar System, the fewer will remain to pose a threat to the Earth, and so such ejection plays a key role in determining the level of impact hazard in the inner Solar System.

Upon examination of Figure 2 (upper panel), it is clear that the time taken for the number of surviving objects to decay to a given value (say half the initial one) is almost a factor of three times longer for simulations without a Jupiter than for those where Jupiter is twice as massive as seen in our own Solar System. Given the assumption that the injection rate of fresh cometary material is the same across all scenarios (an assumption that clearly depends on the initial population of the Oort cloud between scenarios, which is itself reliant on the still debated initial population mechanism for that cloud), it is clear that this would lead to a far greater steady-state population of long period comets for the scenario without a Jupiter, and therefore a greatly enhanced impact threat to the Earth. While it is true that the effects of the galactic tide and passing extra-solar perturbers (such as nearby stars and molecular clouds) will act to strip some of that population away, at the same time as they introduce new members, we believe that this effect would be independent of the mass of Jupiter, and so can safely be ignored.

Figures 3 and 4 add further weight to our conclusions. Figure 3 shows that the ejection of Oort cloud comets on Jupiter crossing orbits (including the Mars and Earth crossers) is more efficient than that for objects which do not cross the giant planet's orbit. Comparison of the plots for the Jupiter, Mars and Earth crossing objects shows that the presence of an increasingly massive "Jupiter" leads to a decrease in the number of Earth- and Mars-crossers – an increasingly massive "Jupiter" acts to remove such objects from the Solar System with an ever increasing efficiency. Figure 4 shows that, even though the mass of "Jupiter" has only a slight effect on the mean orbital period (second panel), it is the increased efficiency with which Oort cloud comets are ejected from the Solar System that results in a decrease in the collision probability as the mass of "Jupiter" increases. The small blue "spike" that can be seen in panels two, three, and four is a result of small number statistics. At the 80 Myr point, a single object had been perturbed to a Jupiter-crossing orbit, meaning that the mean orbital period and mean inverse period were both simply the values for that object. This highlights the risk of drawing conclusions from dangerously small data sets, and we encourage the reader to ignore this datum!

Given these two subsidiary tests, it therefore seems reasonable to conclude that the number of surviving Oort cloud comets as a function of time is a robust proxy for the impact rate resulting from such objects on the Earth.

On the basis of our results, it therefore seems safe to conclude that, of the three populations of potential impactors, it is only in the case of the Oort cloud comets that the planet Jupiter truly is more of a friend than a foe!

Note that we have not considered the effect of the mass of "Jupiter" on the population of the Oort cloud. It is widely thought that the Oort cloud was emplaced early in the Solar System's history by the gravitational scattering of the four giant planets (Morbidelli, 2005). Varying the mass of one of the four would have some effect on the final population, and this could affect the rate at which long period comets enter the inner Solar System, thus modifying our results. To model the early evolution of the Oort cloud population as a function of "Jupiter's" mass (a process which is dependent on a number of uncertain effects such as the local environment in which the Sun formed, and the particular migration and stability history of the giant planets), and then establish the likely change in the rate at which some are scattered inwards, to become long period comets, is a formidable task, a matter for extended future work.

**Conclusions**

As pointed out in Papers I and II, the idea that the planet Jupiter has acted as an impact shield through the Earth's history is one that is entrenched in planetary science, even though little work has been done to examine this idea. In the third of an ongoing series of studies, we have examined the question of Jovian shielding using a test population of particles on orbits representative of the Oort cloud comets, icy bodies that constitute one of three types of potentially hazardous objects (Paper I deals with the asteroids, and Paper II the short-period comets, derived from the Centaurs).

For the Oort cloud comets, the larger the mass of the planet in Jupiter's orbit the greater the rate at which these comets are removed, and therefore the lower the rate of impacts on the Earth. In stark contrast the outcome of Paper I is that *fewer* impacts occur when there is a giant planet of negligible mass in Jupiter's orbit than when Jupiter is present, and from Paper II that there is little difference between the no-Jupiter– Jupiter cases. Both papers show that the impact rate is enhanced by a factor of a few if a giant planet is present in Jupiter's orbit with a mass about 20% that of Jupiter. For the asteroids, we concluded that this is primarily a result of the depth, breadth and location of the $\nu_6$ secular resonance in the main asteroid belt, while for the short-period comets it seems to be due to the interplay between the injection rate of Earth-crossers with the efficiency with which they are removed from the system. Despite the different causes, the similarity between the shapes of the impact distributions is striking. Further work is needed to explore why.

Note that, with impact rates exhibiting a broad peak at about 20% of the mass of Jupiter, our results indicate that exoplanetary systems with giants around the mass of Saturn (30% the mass of Jupiter) could well suffer impact rates on any planets in the inner part of the system far higher than in the Solar System.

Future work will continue the study of the role of Jupiter in limiting or enhancing the impact rate on the Earth by examining the effect of Jovian *location* on the impact fluxes engendered by the three populations. Given the surprising outcome of our work to date, we hesitate to anticipate future outcomes. In particular, given the influence of the $v_6$ resonance on the impact rate experienced from the asteroids as Jupiter's mass is altered, it seems obvious that changes to the location of any Jupiter-like planet (which would in turn cause the web of resonances due to that planet to shift) may make a significant difference to asteroidal-based impacts on terrestrial planets. Additionally, the effect of the eccentricity of Jupiter's orbit needs to be established. This could clearly play a particularly important role in determining the capture and ejection rate of cometary bodies, and merits further study.

Finally, future work will also consider whether the absence of a Jupiter-like body would change the populations of objects which reside in the reservoirs that provide the bulk of the impact hazard in the Solar System, a possible effect ignored in this work.

**Acknowledgements**

This work was carried out with funding from the STFC, and JH and BWJ gratefully acknowledge the financial support given by that body.

FIGURE 1

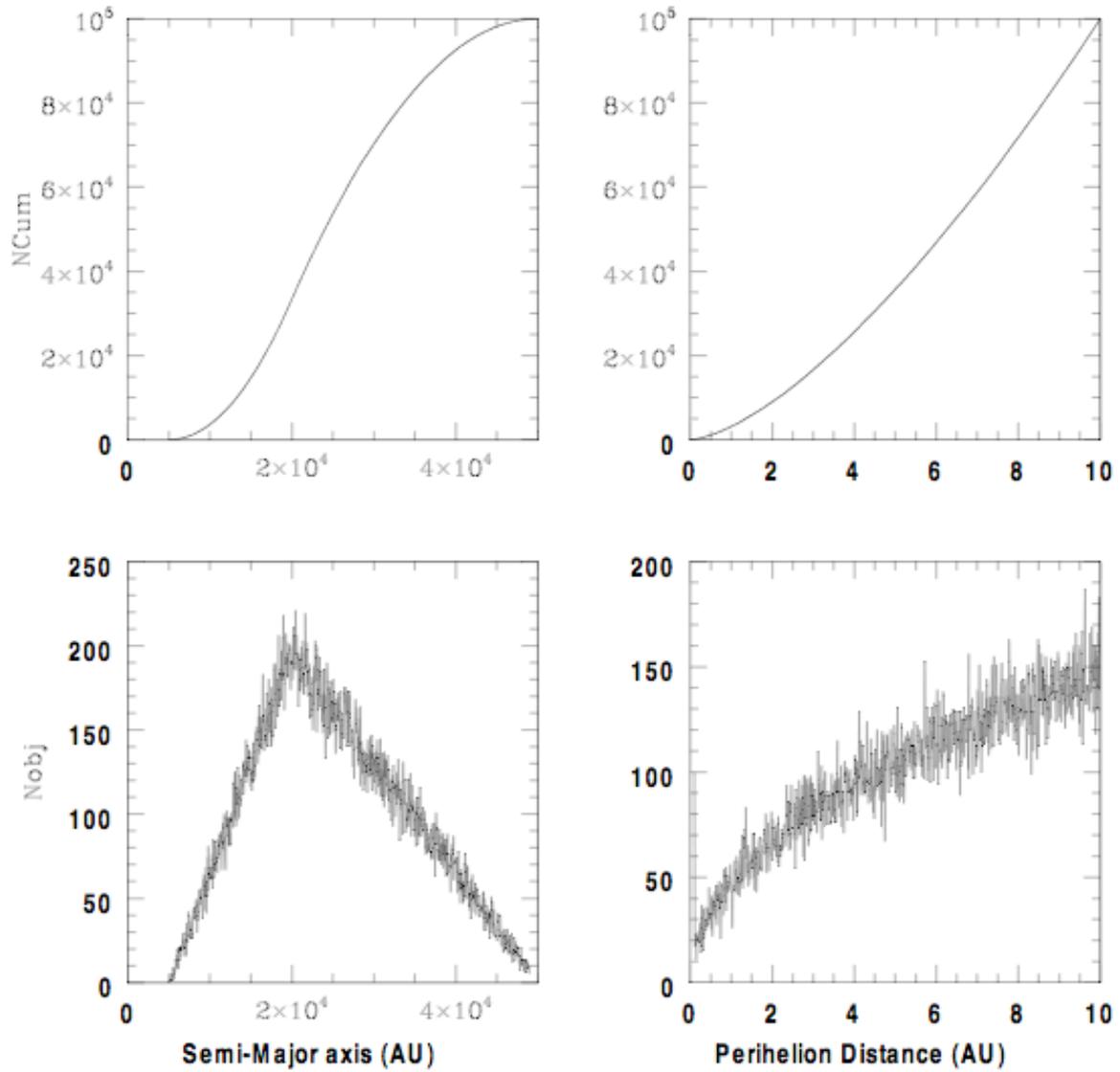

FIGURE 2

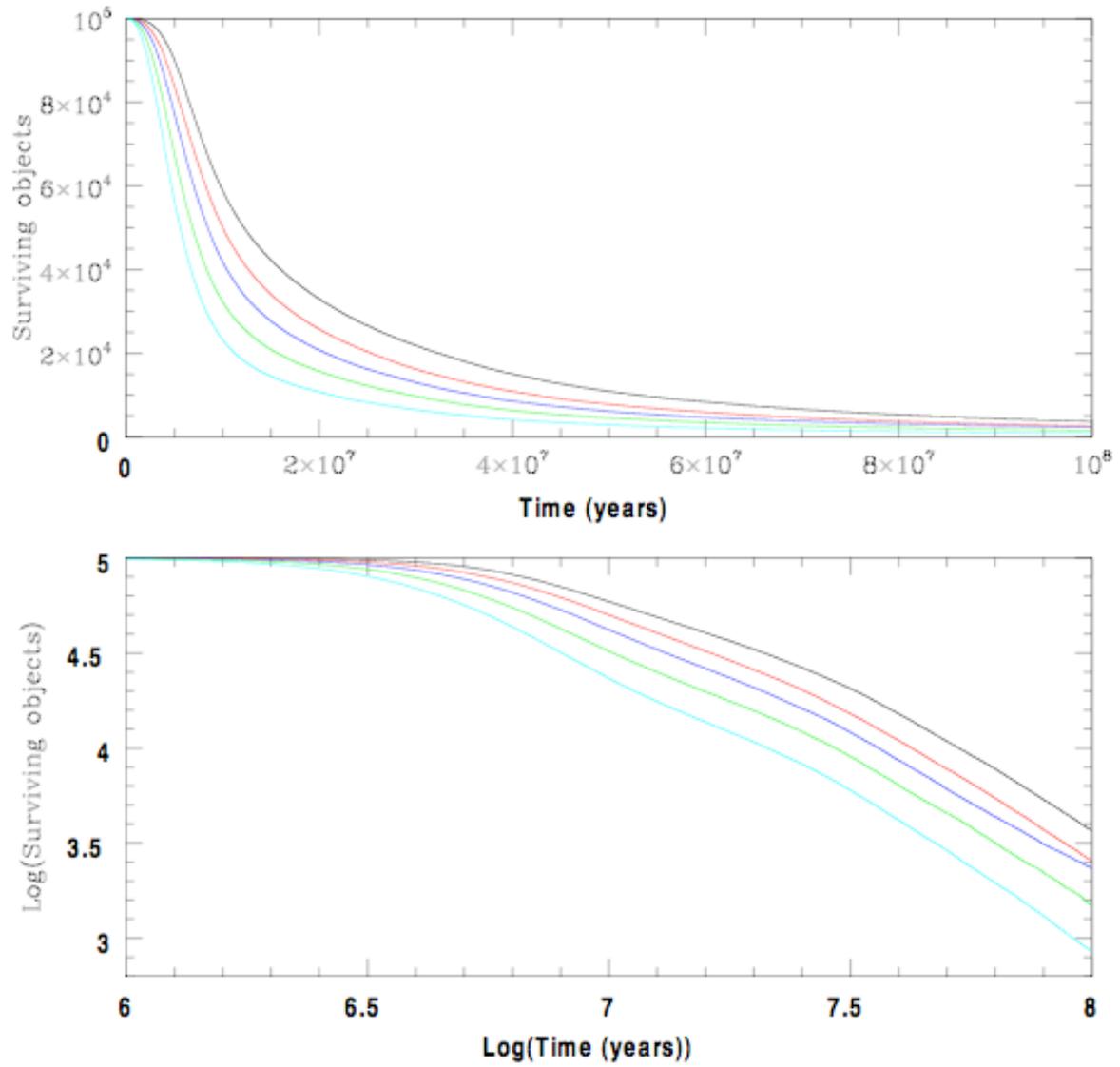

FIGURE 3

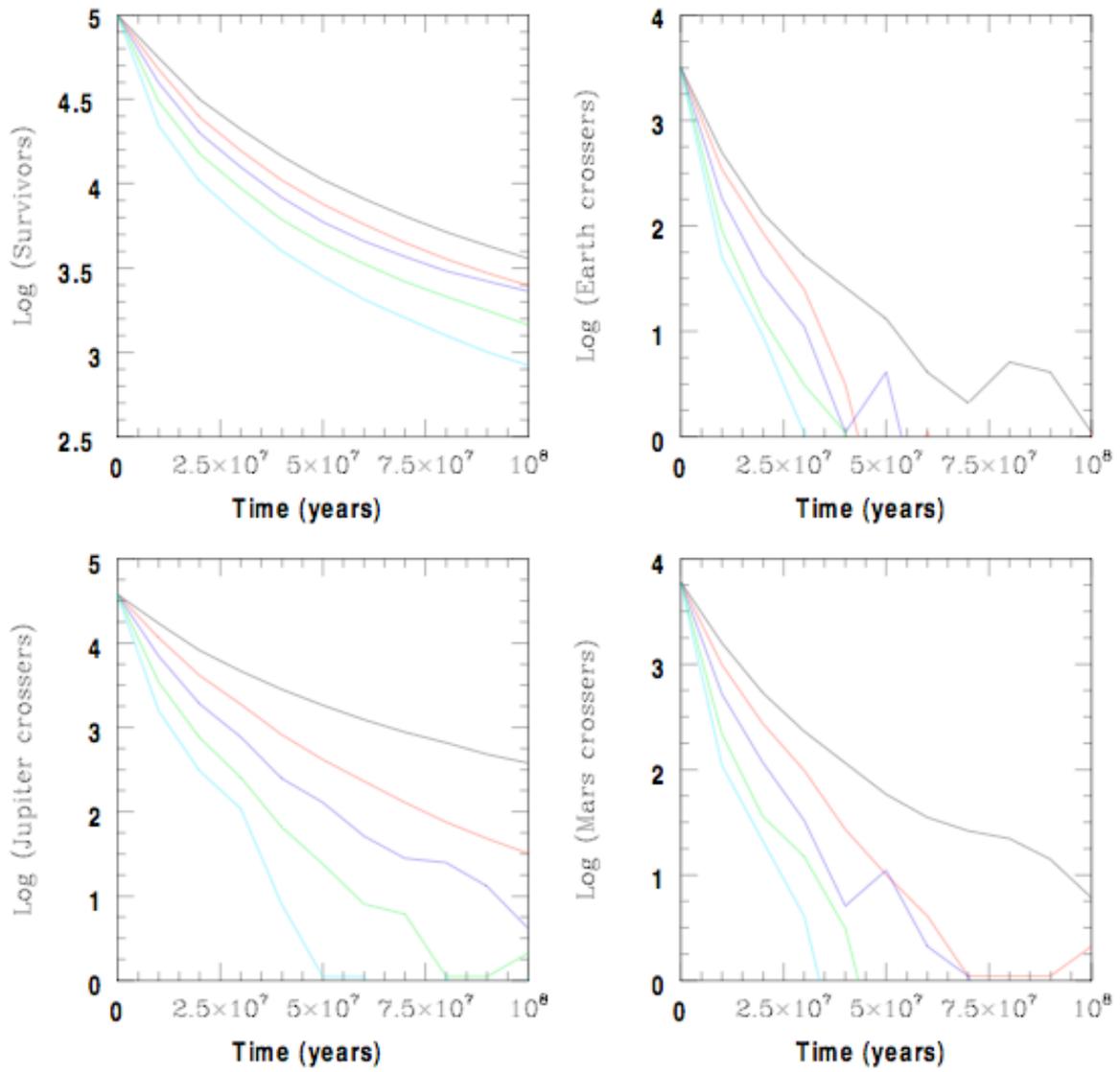

FIGURE 4

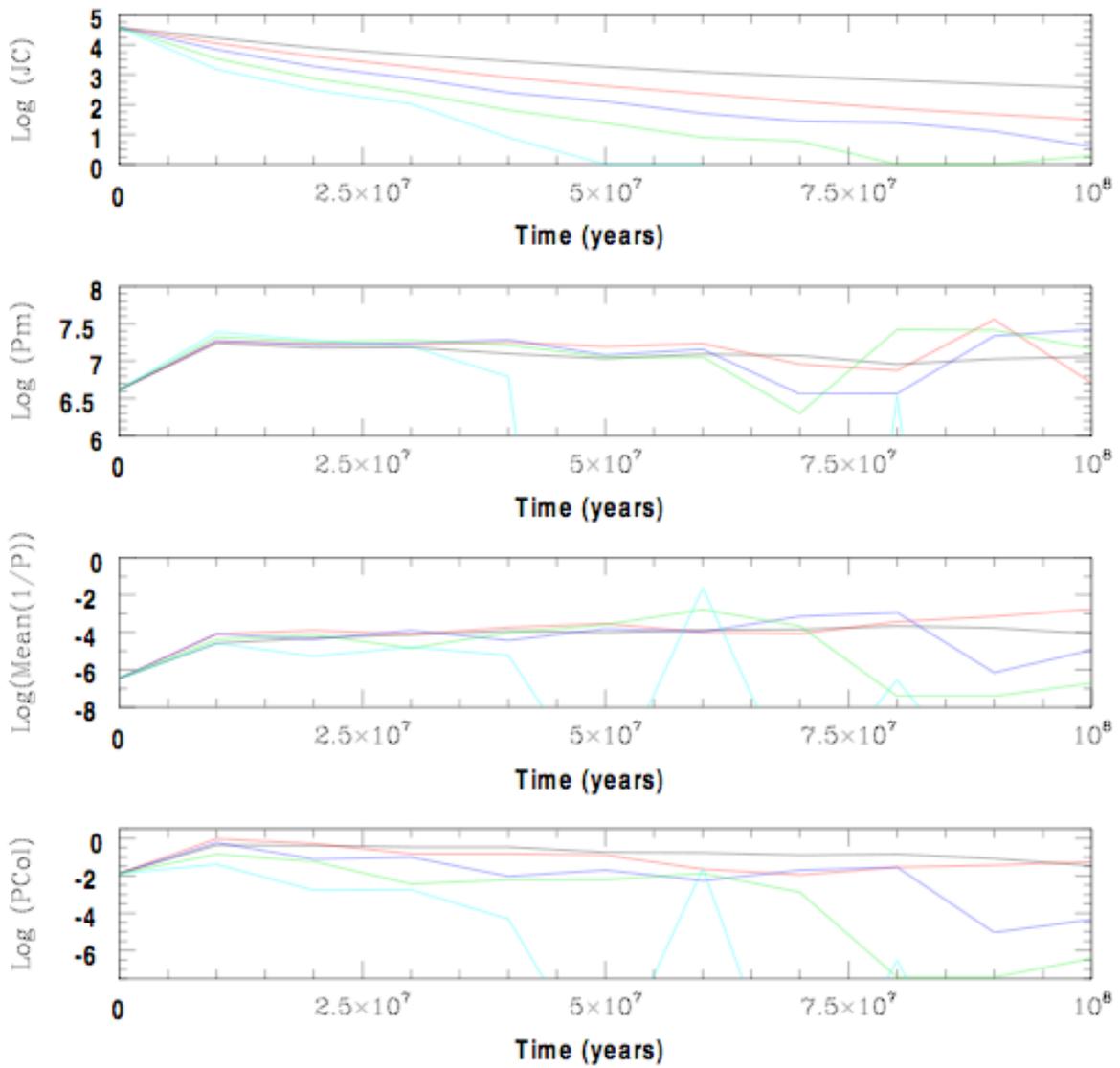